%
%
\def\today{\ifcase\month\or January\or February\or March\or April\or May\or
June\or July\or August\or September\or October\or November\or December\fi
\space\number\day, \number\year}
%
%
\newcount\notenumber

\def\note{\global\advance\notenumber by 1 \footnote{$^{\the\notenumber}$}}
%
%
\newif\ifsectionnumbering
\newcount\eqnumber
\def\cleareqnumber{\eqnumber=0}
\def\numbereq{\global\advance\eqnumber by 1
\ifsectionnumbering\eqno(\the\secnumber.\the\eqnumber)\else\eqno
(\the\eqnumber)\fi}
\def\eqalinno{{\global\advance\eqnumber by 1}
\ifsectionnumbering(\the\secnumber.\the\eqnumber)\else(\the\eqnumber)\fi}
\def\name#1{\ifsectionnumbering\xdef#1{\the\secnumber.\the\eqnumber}
\else\xdef#1{\the\eqnumber}\fi}
\def\nosectionnumbering{\sectionnumberingfalse}
\sectionnumberingtrue
%
%
\newcount\refnumber

\immediate\openout1=refs.tex
\immediate\write1{\noexpand\frenchspacing}
\immediate\write1{\parskip=0pt}
\def\ref#1#2{\global\advance\refnumber by 1%
[\the\refnumber]\xdef#1{\the\refnumber}%
\immediate\write1{\noexpand\item{[#1]}#2}}
\def\tie{\noexpand~}

%
%
\font\twelvebf=cmbx10 scaled \magstep1
\newcount\secnumber

\def\newsection#1.{\ifsectionnumbering\cleareqnumber\else\fi%
	\global\advance\secnumber by 1%
	\bigbreak\medskip\par%
	\line{\twelvebf \the\secnumber. #1.\hfil}\nobreak\smallskip\par}
%
%
%
\def \sqr#1#2{{\vcenter{\vbox{\hrule height.#2pt
	\hbox{\vrule width.#2pt height#1pt \kern#1pt
		\vrule width.#2pt}
		\hrule height.#2pt}}}}

%
%
%
\newdimen\fullhsize
\def\fiddle{\fullhsize=6.5truein \hsize=3.2truein}
\def\fullline{\hbox to\fullhsize}
\def\mkhdline{\vbox to 0pt{\vskip-22.5pt
	\fullline{\vbox to8.5pt{}\the\headline}\vss}\nointerlineskip}
\def\mkftline{\baselineskip=24pt\fullline{\the\footline}}
\let\lr=L \newbox\leftcolumn
\def\twocolumns{\fiddle
	\output={\if L\lr \global\setbox\leftcolumn=\columnbox
		\global\let\lr=R \else \doubleformat \global\let\lr=L\fi
		\ifnum\outputpenalty>-20000 \else\dosupereject\fi}}
\def\doubleformat{\shipout\vbox{\mkhdline
		\fullline{\box\leftcolumn\hfil\columnbox}
		\mkftline} \advancepageno}
\def\columnbox{\leftline{\pagebody}}
\nosectionnumbering
\def\pr#1 {Phys. Rev. {\bf D#1\tie }}
\def\pe#1 {Phys. Rev. {\bf #1\tie}}
\def\pre#1 {Phys. Rep. {\bf #1\tie}}
\def\pl#1 {Phys. Lett. {\bf #1B\tie }}
\def\prl#1 {Phys. Rev. Lett. {\bf #1\tie }}
\def\np#1 {Nucl. Phys. {\bf B#1\tie }}
\def\ap#1 {Ann. Phys. (NY) {\bf #1\tie }}
\def\cmp#1 {Commun. Math. Phys. {\bf #1\tie }}
\def\imp#1 {Int. Jour. Mod. Phys. {\bf A#1\tie }}
\def\mpl#1 {Mod. Phys. Lett. {\bf A#1\tie}}
\def\jmp#1{Jour. Math. Phys. {\bf #1\tie}}
\def\cqg#1{Class. Quant. Grav. {\bf #1\tie}}
\def\tie{\noexpand~}

\def\half{\textstyle{1\over 2}}
\def\quarter{\textstyle{1\over 4}}
\parskip=15pt plus 4pt minus 3pt
\headline{\ifnum \pageno>1\it\hfil Self-Dual
Fields $\ldots$\else \hfil\fi}
\font\title=cmbx10 scaled\magstep1
\font\tit=cmti10 scaled\magstep1
\footline{\ifnum \pageno>1 \hfil \folio \hfil \else
\hfil\fi}
\raggedbottom
\rightline{\vbox{\hbox{RU97-01-B}\hbox{CCNY-HEP-97-1}}}
\vfill
\centerline{\title Symplectic Structures and Self-dual Fields in}
\smallskip
\centerline{\title (4k+2) Dimensions}
\vfill
\centerline{\title Ioannis Giannakis${}^{a}$
and
V. P. Nair${}^{b}$}
\bigskip
\centerline{$^{(a)}$ {\tit Physics Department, Rockefeller
University}}
\centerline{\tit 1230 York Avenue, New York, NY
10021-6399}
\centerline{\rm e-mail: giannak@theory.rockefeller.edu}
\bigskip
\centerline{$^{(b)}${\tit Physics Department, City College
of the City University of New York}}
\centerline{\tit New York, NY10031}
\centerline{\rm e-mail: vpn@ajanta.sci.ccny.cuny.edu}
\vfill
\centerline{\title Abstract}
\bigskip
{\narrower\narrower
We discuss symplectic structures for the chiral boson in $(1+1)$ 
dimensions and the self-dual field in $(4k+2)$ dimensions. Dimensional
reduction of the six-dimensional field on a torus is also considered.}
\vfill\vfill\break

\newsection Introduction.

Self-dual fields, which are described by differential $p$-forms with the
corresponding field strength being self-dual in $(2p+2)$ dimensions, play an
important role in many theories of current interest such as the
six-dimensional and type IIB ten-dimensional supergravities, M-theory and
heterotic strings \ref{\wit}{ E. Witten, IASSNS-HEP-96-101 preprint,
hep-th/9610234; J. Schwarz, CALT-68-2046 preprint, hep-th/9604171;
M. Perry and J. Schwarz, DAMTP-R-96-49 preprint, hep-th/9611065;
P. Argyres and K. Dienes, \pl387 (1996) 727; J. Schwarz,
CALT-68-2091 preprint, hep-th/9701008.}.
It is also well known that there are difficulties in the
construction of manifestly Lorentz-invariant actions for such
fields \ref\sew{N. Marcus and J. Schwarz, \pl115 (1982) 111;
A. Sen and J. Schwarz, \np411 (1994) 35.}. 
Possible ways of circumventing these difficulties have also been recently
explored \ref\ewn{W. Siegel, \np238 (1984) 307; N. Berkovits, \pl388
(1996) 743; IFT-P-039-96 preprint, hep-th/9610134; IFT-P-042-96
preprint, hep-th/9610226; P. Pasti, D. Sorokin and M. Tonin,
\pr52 (1995) 4277; \pl352 (1995) 59.}.
The absence of a manifestly covariant description can be awkward
but does not necessarily hinder the construction of the quantum theory.
The
latter requires a canonical realization of the Poincar\'e algebra with the
equations of motion, viz., the selfduality conditions, being the Heisenberg
equations of motion.
With this motivation, in this letter, we shall explore the canonical structure
for the self-dual fields, directly defining it from the equations of motion.
Such a description can obviously lead to a first-order action. However,
in our approach
the action is more of a derived
quantity. It can also depend on the choice of field variables. One may be
able to choose different fields (e.g., related by duality transformation)
which give different actions.

A simple situation which illustrates these ideas is the free Maxwell theory
with the equations of motion
$$
\partial_0 E_i = \epsilon_{ijk} \partial_j B_k,~~~~~~~~~~~~~~\partial_0
B_i = -\epsilon_{ijk}\partial_j E_k
\numbereq\name{\introa}
$$
with the constraints $\nabla \cdot E=0,~\nabla\cdot B=0$. The
evolution equations involve first derivatives with respect to time and hence
the field variables $(E_i, B_i)$ at a fixed time are sufficient to label
the classical solutions or in other words they describe the phase space
${\cal P}$ of the theory. The canonical structure $\Omega$ is thus a
differential two-form on the space of all fields $(E_i,B_i)$
(at a fixed time). The vector fields which describe the time-evolution and
Poincar\'e transformations of the fields should have the Hamiltonian
property with respect to $\Omega$, i.e., the Lie derivative
of $\Omega$ with respect to the vector field must be zero.
In situations where the first cohomology of the phase space
is trivial this is equivalent to the
contraction of the vector field $\xi$ with $\Omega$, denoted by
$i_{\xi}\Omega$, being the differential of a function on the phase space
$$
i_{\xi}\Omega ~=~-\delta f. \numbereq\name{\introb}
$$
(Here $\delta$ denotes the exterior derivative on ${\cal P}$.) The function
$f$ is the generator of the transformation. For many theories, these
conditions suffice to identify $\Omega$ and the
various generators. ( In what follows we shall impose the
requirement (\introb) rather than the weaker Lie derivative
condition. If the first cohomology of the phase space is nontrivial,
one can have more general solutions to the Lie derivative condition
with associated $\theta$-angles. We shall not consider such situations. )

As an  example, for the
Maxwell theory, time-evolution is described by the vector field
$$
\xi = \int d^3x~ \epsilon_{ijk} \left( \partial_j B_k {\delta \over {\delta
E_i}}-\partial_j E_k{\delta \over {\delta B_i}}\right).
\numbereq\name{\introc}
$$
By writing $\Omega$ in the general form
$$
\Omega = \int_{x,y} \delta E_i(x) \wedge \delta B_j(y) ~M_{ij}(x,y)+
\half \delta E_i(x) \wedge \delta E_j(y) \sigma_{ij}(x,y)+
\half \delta B_i(x)\wedge
\delta B_j(y) \tau_{ij}(x,y) \numbereq\name{\introd}
$$
one can easily see that a solution to the requirement (\introb) is
$$\eqalign{
\Omega &= \int_{x,y} \delta E_i(x) \wedge \delta B_j(y) \epsilon_{ijk}
\partial_k g(x,y)\cr
H&= \half \int_x E_T^2 +B_T^2 \cr}\numbereq\name{\introe}
$$
where $g(x,y)$ is the Coulomb Green's function,
$-\partial^2 g(x,y) =\delta (x-y)$, and $(E_T, B_T)$ denote
the tranverse components of the fields
Defining the potential
$A_i = \int_y \epsilon_{ijk} \partial_k g(x,y) ~B_j(y) $,
we can write $\Omega =\delta \alpha$ with the symplectic potential
$\alpha = \int E_i \delta A_i$.
This leads to the standard first-order action
$$
S= \int E_i \partial_0 A_i ~-H \numbereq\name{\introf}
$$
Alternatively, one can define the dual potential
${\tilde A}_i= -\int \epsilon_{ijk}\partial_k g(x,y) ~E_j(y)$ with the
symplectic potential ${\tilde \alpha} = \int B_i \delta {\tilde A}_i$;
this gives the dual description. ( See in this connection
\ref{\rja}{R. Jackiw in {\it Diverse Topics in Theoretical
and Mathematical Physics}, World Scientific, Singapore, (1995).}.)
 
In what follows, we explore a similar approach for self-dual fields in
$(4 k+2)$ dimensions. Our approach, then, would be as follows. We start
with the equations of motion of the underlying theory.
Some of these describe the time-evolution of
fields, some are constraints on the fields at a given time.
By writing the equations of motion in first-order form one can identify
a suitable set of variables as coordinates for the phase space
of the theory. Subsequently we introduce a
symplectic structure on the phase space which should be
a closed two-form. The time-evolution equations of
motion must correspond to a Hamiltonian vector field for this
symplectic two-form. ( More generally all Poincar\'e transformations
must correspond to Hamiltonian vector fields.) Strictly
speaking, the symplectic
two-form should be nondegenerate but very often it would be more convenient
to start with a two-form which has degenerate directions. The constraints
on the fields must correspond precisely to the degenerate directions or
to gauge symmetries. Starting with the equations
of motion and using these requirements, in many situations one
is able to obtain a canonical formulation and a first-order
action for the theory. It is also possible to construct functions on
the phase space that generate the spacetime symmetries of the
theory. Quantization is then fairly straightforward.
(A similar approach has been used for analyzing the interactions
of anyons in \ref\pol{ C. Chou, V. P. Nair and A. Polychronakos,
\pl304 (1993) 105.}.)

Most of our discussion will be for the chiral
boson in $(1 + 1)$ dimensions and the self-dual field in
six dimensions. We shall also discuss
the dimensional reduction of latter to four dimensions with the resulting
dual symmetric electrodynamics \ref\verli{ E. Verlinde, \np455 (1995)
211; D. Berman, CERN-TH-96-366 preprint, hep-th/9612191.}.
The supersymmetric version of the
six-dimensional case and the general
$(4k+2)$-dimensional case are commented upon. 

\newsection Chiral boson in $(1+1)$ dimensions.

This is described by a scalar field $\phi(x, t)$ which obeys the equation
of motion
$$
\partial_{0}{\phi(x, t)}-\partial_{1}{\phi(x, t)}=0
\numbereq\name{\equrth}
$$
where $x^0=t$ and  $x^1=x$. The equation of motion
being first-order in
time derivatives, the classical trajectories can be labelled by the initial
value of $\phi(x,t)$. Thus the phase space can be described by
field configurations $\phi(x, t)$ at a fixed time $t$.
The symplectic two-form is thus given by
$$
\Omega={\half}{\int_{x,y}}M(x, y, \phi){\delta{\phi(x)}}
\wedge{\delta{\phi(y)}}
\numbereq\name{\eqaxui}
$$
where we denote exterior differentiation on the space of $\phi$'s
by $\delta$. Evidently from the antisymmetry of the exterior product,
$M(x, y)=-M(y, x)$.

The time-evolution of $\phi(x, t)$ as given in (\equrth) corresponds
to the vector field
$$
\xi=\int_{x}{\partial_{1}}{\phi}{\delta \over {\delta{\phi}}}.
\numbereq\name{\eqrantos}
$$
The requirement that $\xi$ be a Hamiltonian vector field for $\Omega$
leads to
$$
i_{\xi}{\Omega}=\int_{x}{\partial_{x}}{\phi}(x)M(x, y, \phi)
{\delta{\phi(y)}}=-\int_{x}{\phi}(x){\partial_{x}}M(x, y, \phi)
{\delta{\phi(y)}}=-{\delta}H.
\numbereq\name{\eqvares}
$$
The simplest solution to this is given by $M(x, y, \phi)$ which is
independent of $\phi$. In this case we can satisfy (\eqvares)
by the choice
$$
{\partial_{x}}M(x, y)=\delta(x, y).
\numbereq\name{\eqkarata}
$$
This gives
$$
H={\half}{\int_{x}}{\phi}^2(x).
\numbereq\name{\eqgewrga}
$$
Given (\eqkarata), we find $i_{V}{\Omega}=-{\delta{\phi}}$
where $V(x)$, the vector field corresponding to $\phi$, is
given by $-{\partial_{x}}{\delta \over {\delta{\phi}}}$. The
fundamental Poisson bracket is thus
$$
\big [ \phi(x), \phi(y)\big ]=i_{V(x)}i_{V(y)}{\Omega}=
\partial_{x}{\delta}(x-y).
\numbereq\name{\eqntambi}
$$
The symplectic potential $\alpha$ defined by $\Omega=
{\delta{\alpha}}$ is given by
$$
\alpha={\half}{\int_{x,y}}{\phi}(x)M(x, y){\delta{\phi(y)}}.
\numbereq\name{\eqsamba}
$$
Once we have the symplectic potential, a first-order action for the theory
can be constructed. In the present case it is given by
$$
S={\half}{\int_{x,y,t}}{\phi}(x)M(x, y){\dot{\phi}}(y)
-{\half}{\int_{x,t}}{\phi}^{2}(x).
\numbereq\name{\eqsfakia}
$$
(The overdot on the field $\phi$ denotes the derivative with respect to time.)

Linear momentum $P$ corresponds to the vector field
$\zeta=\int{\partial_{x}}{\phi}{\delta \over {\delta{\phi}}}$;
evidently $H=P$. Consider Lorentz transformations. If $\phi$
is a scalar, we have, infinitesimally,
$$
\delta{\phi(x, t)}=\upsilon(t{\partial_{x}}{\phi}+x{\partial_{t}}
{\phi})=\upsilon(t+x){\partial_{x}}{\phi}
\numbereq\name{\eqkarasa}
$$
which corresponds to a vector field
$$
{\tilde v}_{L}=\int(t+x){\partial_{x}}{\phi}{\delta \over {\delta{\phi}}}.
\numbereq\name{\eqpasalh}
$$
The contraction of this vector field with $\Omega$ is given by
$$
i_{{\tilde v}_{L}}{\Omega}=-t{\delta}H-{\int_{x}}x{\delta}T^{00}-
{\int_{x,y}}{\phi}(x)M(x, y){\delta{\phi(y)}}=
-{\delta}(tH+{\int_{x}}x T^{00})-2{\alpha}
\numbereq\name{\eqkarapial}
$$
where $T^{00}={\half}{\phi}^2$ is the energy density. We cannot
write $\alpha$ as $\delta f$ since $\delta{\alpha}=\Omega\ne 0$.
Thus ${\tilde v}_{L}$ is not a Hamiltonian vector field. In order
to achieve this we must
modify the transformation property of $\phi$, Eq.(\eqkarasa).
A suitable modification is given by
$$
{v}_{L}=\int(t+x){\partial_{x}}{\phi}{\delta \over {\delta{\phi}}}
+{\int}{\phi}{\delta \over {\delta{\phi}}}, \qquad
i_{{ v}_{L}}{\Omega}=-{\delta}(tH+{\int_{x}}x T^{00}).
\numbereq\name{\eqgeorgev}
$$
The generator of Lorentz transformations is thus
$$
L=tH+{\half}{\int_{x}}x{\phi}^{2}(x).
\numbereq\name{\eqamanat}
$$
The full Poincar\'e algebra is easily checked for $H, P, L$.
The transformation property of $\phi$
$$
\big [ L, \phi(x) \big ]=-(t+x){\partial_{x}}{\phi}(x)-{\phi}(x)
\numbereq\name{\eqivic}
$$
shows that $\phi$ has nontrivial spin.

As far as time-evolution is concerned, it is possible to have more general
solutions to (\eqvares). For example we can take
${\partial_{x}}M(x, y)=G(x, y)$, $H={\half}{\int_{x, y}}{\phi}(x)
G(x, y){\phi}(y)$ where $G(x, y)$ is some symmetric function of $x, y$.
For the Lorentz transformation we then find
$$
i_{{ v}_{L}}{\Omega}=-{\delta}(tH+\int_{x}xT^{00})-
{\half}{\int_{x, y}}(x-y){\phi}(x)G(x, y){\delta{\phi(y)}}.
\numbereq\name{\eqkaklama}
$$
For the Poincar\'e algebra to be satisfied, the last term must vanish.
Since this must be true for any field we must have $(x-y)G(x, y)=0$,
with the solution that $G(x, y)$ is proportional to ${\delta}(x- y)$.
For the chiral boson in $(1+1)$-dimensions this basically leads to the
action (\eqsfakia) and the Poisson bracket (\eqntambi) with
$M(x, y)$ given by (\eqkarata). What we have obtained is the
Floreanini-Jackiw description of the chiral boson \ref\jac{R. Floreanini
and R. Jackiw, \prl59 (1987) 1873.}.

\newsection Self-dual field in six dimensions.

In order to study the characteristics of self-dual theories
where the antisymmetric tensors are gauge fields, it is sufficient
to consider the six-dimensional case. Generalization
to higher dimensions is then straightforward.
The gauge field in six dimensions is given by a two-form potential
$C$. Its field strength $H=dC$ is self-dual and it is
invariant under the gauge transformations $C \to C+dN$, where
$N$ is a one-form. The equations of motion, viz., the selfduality
of $H$, are given by
$H=^* H$, $~^*$ denoting the Hodge dual. In components
this reads
$$
{\partial_{0}}C_{\mu\nu}+{\partial_{\nu}}C_{0\mu}
+{\partial_{\mu}}C_{\nu0}={1\over 3!}{\epsilon_{0\mu\nu
\alpha\beta\gamma}} \big ({\partial_{\alpha}}C_{\beta\gamma}
+{\partial_{\gamma}}C_{\alpha\beta}+{\partial_{\beta}}
C_{\gamma\alpha} \big )
\numbereq\name{\eqalexa}
$$
with $(\mu, \nu, \cdots=1,2, \cdots 5)$. We can use the gauge freedom
to set $C_{0\mu}=0$ and simplify the
equations of motion. This will be assumed in what follows. Although we
have imposed a gauge condition, there is
still the residual freedom of time-independent gauge transformations,
$C_{\mu\nu} \to C_{\mu\nu}+{\partial_{\mu}}N_{\nu}
-{\partial_{\nu}}N_{\mu}$. The equations of motion are first order in
time and hence solutions of (\eqalexa) may be labelled by the set of
configurations $C_{\mu\nu}(x)$ at a fixed time, denoted
by $\hat {\cal P}$. There
is still redundancy in the parametrization because of the freedom
of time-independent gauge transformations. The true phase space, ${\cal P}$,
is obtained from $\hat {\cal P}$ by the identification of configurations which
only differ by a gauge transformation. We can however start
with a symplectic two-form $\Omega$ on $\hat {\cal P}$. Such
an $\Omega$ should have zero components
along the gauge directions of the phase space. Indeed the
vector field $V_g$ generating the gauge transformation, viz.,
$$
V_g={\int} \big ( {\partial_{\mu}}N_{\nu}
-{\partial_{\nu}}N_{\mu} \big ) {\delta \over {{\delta}C_{\mu\nu}}}
\numbereq\name{\eqsapanhs}
$$
should be a zero mode of $\Omega$, i.e., $i_{V_{g}}{\Omega}=0$.
For $\Omega$, we can assume the general form
$$
\Omega={\half}{\int_{x,y}}M^{\mu\nu\kappa\lambda}
(x, y, C)~{\delta}C_{\mu\nu}(x)
\wedge{\delta}C_{\kappa\lambda}(y)
\numbereq\name{\eqtroupk}
$$
where $M^{\mu\nu\kappa\lambda}(x, y)=-
M^{\kappa\lambda\mu\nu}(y, x)$. The requirement of
$i_{V_{g}}{\Omega}=0$ gives ${\partial_{\mu}}M^{\mu\nu\kappa\lambda}
(x, y)=0$, a solution to which may be taken as
$$
M^{\mu\nu\kappa\lambda}(x, y)={\epsilon_{\mu\nu
\kappa\lambda\rho}}{\partial^{\rho}}F(x, y).
\numbereq\name{\eqioanni}
$$
Since $\Omega$ has zero modes, it is not invertible. In order to have
an invertible $\Omega$ we have to impose a gauge-fixing condition.
Upon gauge-fixing and inverting we get the Dirac brackets. These have the
property that ${\big [ {\phi}, f \big ]}_D=0$ for any observable $\phi$
where $f$ is the gauge-fixing constraint. In our case we can construct
the basic bracket relations without going through this formal procedure.
With $M^{\mu\nu\kappa\lambda}(x, y)$ as given by (\eqioanni) we can easily
check that the vector field
$\rho^{\mu\nu}\equiv{\delta \over {{\delta}C_{\mu\nu}}}$ is a
Hamiltonian vector field if $F(x, y)$ is independent of $C_{\mu\nu}$.
The corresponding function on phase space is given by
$$
P^{\mu\nu}(x)=-{\int_{z}}{\epsilon_{\mu\nu
\kappa\lambda\rho}}C_{\kappa\lambda}(z){\partial^{\rho}}F(x, z)
={\int_{z}}{\epsilon_{\mu\nu
\kappa\lambda\rho}}{\partial^{\rho}}C_{\kappa\lambda}(z)F(x, z).
\numbereq\name{\eqeleu0e}
$$
The fundamental bracket relations are thus
$$
\big [ P^{\mu\nu}(x), P^{\kappa\lambda}(y) \big ]=
\big ( i_{\rho^{\mu\nu}} i_{\rho^{\kappa\lambda}}
{\Omega} \big )=-{\epsilon_{\mu\nu
\kappa\lambda\rho}}{\partial^{\rho}}F(x, y).
\numbereq\name{\eqstrako}
$$
Notice that $P^{\mu\nu}(x)$ is gauge-invariant. In solving (\eqstrako)
to obtain the brackets for $C_{\mu\nu}$'s, one has to choose a gauge-fixing
condition. From (\eqalexa) we see that time-evolution is generated by
$$
\xi={\half}{\int}{\epsilon_{\mu\nu
\kappa\lambda\rho}}{\partial^{\rho}}C_{\kappa\lambda}
{\delta \over {{\delta}C_{\mu\nu}}}.
\numbereq\name{\eqgkonias}
$$
We then find
$$\eqalign{
i_{\xi}{\Omega}&=-{\int_{x,y}}C_{\beta\gamma}(x) {\delta}C_{\rho\sigma}(y)
L^{\beta\gamma,\rho\sigma}F(x, y)\cr
L^{\beta\gamma,\rho\sigma}&=\big [ {\delta}^{\rho\sigma}_{\beta\gamma}
{\partial^2}-{\delta}^{\rho\sigma}_{\alpha\gamma}{\partial_{\alpha}}
{\partial_{\beta}}+{\delta}^{\rho\sigma}_{\alpha\beta}{\partial_{\alpha}}
{\partial_{\gamma}} \big ] \cr}
\numbereq\name{\eqpantelic}
$$
As it stands, $\xi$ is not a Hamiltonian vector field, due to the
${\partial_{\alpha}}
{\partial_{\beta}}$- and ${\partial_{\alpha}}
{\partial_{\gamma}}$-terms in the above expression.
Also the operator $L^{\beta\gamma,\rho\sigma}$
has zero modes and is not invertible; this is a reflection of the residual
gauge invariance. A convenient gauge-fixing
condition is given by ${\partial^{\mu}}C_{\mu\nu}(x)=0$. As can be
easily seen, this condition is preserved in time by the evolution equations
(\eqalexa) (with $C_{0\mu}(x)=0$ of course). In this case (\eqpantelic)
simplifies as
$$
i_{\xi}{\Omega}=-2{\int_{x,y}}C_{\rho\sigma}(x)
{\delta}C_{\rho\sigma}(y)~{\partial^2}F(x, y)=-{\delta}H
\numbereq\name{\eqkalatz}
$$
with $H={\int_{x,y}}C_{\rho\sigma}(x){\partial^2}F(x, y)C_{\rho\sigma}(y)$.
Thus $\xi$ is indeed a Hamiltonian vector field and $H$ is the Hamiltonian.
With the gauge-fixing condition ${\partial^{\mu}}C_{\mu\nu}(x)=0$ we can
solve (\eqstrako) to obtain the bracket relations for $C_{\mu\nu}(x)$ as
$$
\big [ C_{\mu\nu}(x), C_{\alpha\beta}(y) \big ]={\quarter}
{\epsilon_{\mu\nu
\alpha\beta\gamma}}{\partial^{\gamma}}{\Delta}(x, y)
\numbereq\name{\eqtarlac}
$$
where $\Delta(x, y)$ is the inverse of ${\partial^2}F(x, y)$. A simple choice
which is consistent with Lorentz symmetry is to take
$$
{\partial^2}F(x, y)=\delta(x-y), \qquad \Delta(x, y)=\delta(x-y).
\numbereq\name{\eqsigalas}
$$
In this case we have for the energy density $T^{00}(x)=C_{\alpha\beta}(x)
C_{\alpha\beta}(x)$ with $H={\int_{x}}T^{00}(x)$ and also (\eqtarlac) becomes
$$
\big [ C_{\mu\nu}(x), C_{\alpha\beta}(y) \big ]={\quarter}
{\epsilon_{\mu\nu
\alpha\beta\gamma}}{\partial^{\gamma}}{\delta}(x, y)
\numbereq\name{\eqtomic}
$$
The relation (\eqtomic) is identical to the bracket relation given
by Henneaux and Teitelboim in \ref\teitl{ M. Henneaux and
C. Teitelboim, \pl206 (1988) 650.}. Notice also that (\eqtomic)
gives $ \big [ {\partial^{\mu}}C_{\mu\nu}(x), C_{\alpha\beta}(y) \big ]=0$.

Using the expression for $T^{00}(x)$ we can obtain some of the other
components of the energy-momentum tensor. For example, the
bracket relation
$$
\big [ T^{00}(x), T^{00}(y) \big ]= \big ( T^{0\alpha}(x)
+ T^{0\alpha}(y) \big ) {\partial_{\alpha}}{\delta}(x-y)
\numbereq\name{\eqbakats}
$$
which holds for a Poincar\'e-invariant theory can be used, along
with eqs. (\eqsigalas) and (\eqtomic), to identify $T^{0\alpha}(x)$ as
$$
T^{0\alpha}(x)={\half}{\epsilon^{\alpha\mu\nu\beta\gamma}}
C_{\mu\nu}(x)C_{\beta\gamma}(x).
\numbereq\name{\eqrivers}
$$
The linear momentum is given by $P^{\alpha}={\int_{x}}T^{0\alpha}(x)$
and obeys the relation
$$
\big [ P^{\alpha}, C_{\mu\nu}(x) \big ]=-{\partial^{\alpha}}
C_{\mu\nu}(x)
\numbereq\name{\eqpapan}
$$
up to the condition ${\partial^{\mu}}C_{\mu\nu}=0$.

We now consider Lorentz transformations. Naively the orbital
part of the Lorentz boosts corresponds to a vector field
$$
{\tilde V}_L={\int_{x}} \big [ x^{0}v^{j}{\partial_{j}}C_{\mu\nu}(x)
+{\half}v^{j}x^{j}{\epsilon_{\mu\nu
\alpha\beta\gamma}}{\partial_{\alpha}}C_{\beta\gamma}(x) \big ]
{\delta \over {{\delta}C_{\mu\nu}}}.
\numbereq\name{\eqfasou}
$$
It is easily seen that ${\tilde V}_L$ is not a Hamiltonian
vector field, i.e., $i_{{\tilde V}_L}\Omega \ne -{\delta}f$ for some $f$.
One has to modify the vector field to
$$
V_L={\int_{x}} \big [ x^{0}v^{j}{\partial_{j}}C_{\mu\nu}(x)
+{\half}v^{j}x^{j}{\epsilon_{\mu\nu
\alpha\beta\gamma}}{\partial_{\alpha}}C_{\beta\gamma}(x)
+v^{\kappa}{\epsilon_{\mu\nu
\rho\sigma\kappa}}C_{\rho\sigma}(x) \big ]
{\delta \over {{\delta}C_{\mu\nu}}}.
\numbereq\name{\eqfasoula}
$$
In this case $i_{V_L}= -{\delta}(v^{\mu}L_{\mu})$ with
$$
L^{\mu}=x^{0}P^{\mu}-{\int_{x}}x^{\mu}T^{00}(x).
\numbereq\name{\eqfa}
$$
The anomalous term, viz., $v^{\kappa}{\epsilon_{\mu\nu
\rho\sigma\kappa}}C_{\rho\sigma}(x)$ in (\eqfasoula) is as
expected. This is the ``operator gauge transformation'' part which
arises because we have used the non-invariant gauge condition
$C_{0\mu}(x)=0$ ( For a discussion on the operator
gauge transformation which reestablishes
the Coulomb gauge in the new Lorentz frame in
electrodynamics see \ref\zou{ B. Zumino,
\jmp1 (1960) 1.}). From the relation $\big [ L^{\mu}, L^{\nu} \big ]=
M^{\mu\nu}$, we identify the spatial angular momentum as
$$
M^{\mu\nu}={\int_{x}} \big ( x^{\mu}T^{0\nu}(x)-x^{\nu}T^{0\mu}(x)
\big ).
\numbereq\name{\eqgcx}
$$
This has the bracket relation
$$
\eqalign{
\big [ M^{\mu\nu}, C^{\alpha\beta}(x) \big ]= -\big ( x^{\mu}{\partial^{\nu}}-
x^{\nu}{\partial^{\mu}} \big )C^{\alpha\beta}(x)+ \big ( {\delta}^{\nu\alpha}
C^{\mu\beta}(x)-{\delta}^{\nu\beta}
C^{\mu\alpha}(x)&+{\delta}^{\mu\beta}
C^{\nu\alpha}(x)\cr
&-{\delta}^{\mu\alpha}
C^{\nu\beta}(x) \big )\cr}
\numbereq\name{\eqgdsk}
$$
which shows the infinitesimal change of $C^{\alpha \beta}$
corresponding to the orbital and spin transformations.
Similarly we can construct the generators of dilatations and
special conformal transformations. Dilatations correspond to
a Hamiltonian vector field
$$
\zeta={\int_{x}} \bigl[~\partial_\gamma (x^\mu C^{\gamma \nu}+x^\nu
C^{\mu \gamma}+x^\gamma C^{\nu\mu})~+{\half}x^{0}\epsilon_{0\mu\nu
\alpha\beta\gamma}\partial_\alpha C_{\beta \gamma}\bigr]
{\delta \over {{\delta}C_{\mu\nu}}}
\numbereq\name{\eqlaios}
$$
such that $i_{\zeta}{\Omega}=-{\delta}D$, with the generator of dilatations
given by
$$
D={\int_{x}} \big ( x^{0}T^{00}(x)+x^{\alpha}T^{0\alpha}(x)
\big ).
\numbereq\name{\eqpolyzo}
$$
The bracket relation
$$
\big [ D, C_{\alpha\beta}(x) \big ]= -3 C_{\alpha\beta}(x) -
x^{\kappa}{\partial_{\kappa}}C_{\alpha\beta}(x)
-x^{0}{\partial_{0}}C_{\alpha\beta}(x).
\numbereq\name{\eqdoriad}
$$
gives the transformation of $C^{\alpha \beta}$ under dilatations.
Finally we consider special conformal transformations. Naively they
correspond to a vector field $\tilde\zeta$
$$
\tilde\zeta={\int_{x}} \big (2k^{\lambda}x^{\nu}C_{\mu\lambda}
-2k^{\nu}x^{\lambda}C_{\mu\lambda}+2k^{\lambda}x^{\mu}C_{\lambda\nu}
-2k^{\mu}x^{\lambda}C_{\lambda\nu}-6k^{A}x^{A}
C_{\mu\nu}+{\delta}x^{A}{\partial_{A}}
C_{\mu\nu}\big ) {\delta \over {{\delta}C_{\mu\nu}}}
\numbereq\name{\eqdoridma}
$$
with ${\delta}x^{A}=k^{B}(2x^{B}x^{A}-x^{2}{\delta^{AB}})$
and $A, B=0, \cdots ,5$.
As with Lorentz transformations, for this vector field to have the
Hamiltonian property, we need to modify the
transformation properties of $C_{\mu\nu}$ under special
conformal transformations. More specifically we need to add to
${\tilde \zeta}$ the term
$$
{\tilde\zeta}'={\int_{x}}\big (
{\epsilon^{\mu\nu\lambda\rho\sigma}} k^{\lambda}x^{0}C_{\rho\sigma}
-{\epsilon^{\mu\nu\lambda\rho\sigma}}k^{0}x^{\lambda}
C_{\rho\sigma} \big ){\delta \over {{\delta}C_{\mu\nu}}}
\numbereq\name{\eqgiand}
$$
The vector field $\zeta ={\tilde \zeta}+{\tilde \zeta}'$ is Hamiltonian,
i.e., $i_{{\tilde\zeta}+{\tilde\zeta}'}{\Omega}=
-{\delta}(k^{A}K^{A})$ with
$$
K^{A}={\int_{x}} \big ( x^{2}{\delta^{AB}}-2x^{A}x^{B}\big )T^{0B}(x)
\numbereq\name{\eqddnd}
$$
The transformation
properties of the gauge field $C_{\mu\nu}$ under special
conformal transformations also reveal an anomalous term, viz.,
${\epsilon^{\mu\nu\lambda\rho\sigma}} k^{\lambda}x^{0}C_{\rho\sigma}
-{\epsilon^{\mu\nu\lambda\rho\sigma}}k^{0}x^{\lambda}
C_{\rho\sigma}$ which reestablishes the gauge condition
$\partial_{\mu}C_{\mu\nu}(x)=0$ after the transformation.

The discussion upto this point shows that the symplectic structure $\Omega$
of eqs.(\eqtroupk, \eqioanni) with $F$ given by eq.(\eqsigalas) gives a
Poincar\'e-invariant (actually conformally invariant) description of the
selfdual field. We can now consider an action for this field.
The symplectic potential $\alpha$ for the two-form $\Omega$ is given by
$$
\alpha={\half}{\int_{x, y}}C_{\mu\nu}(x){\delta}C_{\alpha\beta}(y)
{\epsilon^{\mu\nu\alpha\beta\gamma}}{\partial_{\gamma}}F(x, y).
\numbereq\name{\eqwelp}
$$
A first order action for the self-dual field is thus
$$
S={\half}{\int_{x, y}}C_{\mu\nu}(x){\dot C}_{\alpha\beta}(y)
{\epsilon^{\mu\nu\alpha\beta\gamma}}{\partial_{\gamma}}F(x, y)
-{\int_{x}}C_{\mu\nu}(x)C_{\mu\nu}(x).
\numbereq\name{\eqwaxops}
$$

Our analysis goes over in a straightforward way to $(4k+2)$ dimensions.
The fields in this case are described by a $2k$-form with components
$C_{\mu_1 \mu_2 ...\mu_{2k}},~\mu_i =1,2,...,(2k+1)$ and
$C_{0 \mu_1 ...}=0$ as before.
$\Omega$ is given by
$$
\Omega =\half \int
\epsilon_{\mu_1 \mu_2 ...\mu_{2k} \nu_1 \nu_2 ...\nu_{2k} \alpha}
\delta C_{\mu_1 \mu_2 ...\mu_{2k}}(x)\wedge
\delta C_{\nu_1 \nu_2 ...\nu_{2k}}(y)
\partial_\alpha F(x,y)
\numbereq\name {\highera}
$$
with $\partial^2 F(x,y)=\delta (x-y)$. We also have
$$
T^{00}= \half (2k)! ~C_{\mu_1 \mu_2 ...\mu_{2k}}(x)
C_{\mu_1 \mu_2 ...\mu_{2k}}(x)
\numbereq\name{\higherb}
$$
The basic Poisson brackets are
$$
\left[ C_{\mu_1 \mu_2 ...\mu_{2k}}(x), C_{\nu_1 \nu_2 ...\nu_{2k}}(y)\right]
= {1\over {((2k)!)^2}}
\epsilon_{\mu_1 \mu_2 ...\mu_{2k} \nu_1 \nu_2 ...\nu_{2k} \alpha}
\partial_\alpha \delta (x-y)
\numbereq\name{\higherc}
$$

\newsection The supersymmetric self-dual field.

In many situations, the supersymmetric version of the self-dual
field is of interest. This has been studied before and here we simply
point out that supersymmetrization does not lead to
new features within our approach.
For supersymmetrization of the six-dimensional case
with Euclidean signature, we need
complex $C_{\mu\nu}$ since self-duality requires complex fields
\ref\rom{ L. Romans, \np276 (1986) 71;
A. Sagnotti, \pl294 (1992) 196; E. Bergshoeff, E. Sezgin
and E. Sokatchev, \cqg{13} (1996) 2875.}.
Further we have two Weyl spinors $\psi^{A}_{i}$, $A=1,2, i=1,2,3,4$
which transform on the $i$-index as the fundamental representation
of $SU(4)$ (i.e., spinor of $SO(6)$). We also have a complex
scalar $\phi$. With Minkowski signature, one can impose a
symplectic Majorana condition on the $\psi$'s and make
$C_{\mu\nu}, \phi$ real. The supersymmetry transformations
(with Euclidean signature ) are of the form
$$
\eqalign{
\delta C_{\mu\nu}&={\overline{\psi}}^{A}{\gamma_{\mu\nu}}{\eta_{A}}
+{\eta^{*}_{A}}{\gamma_{\mu\nu}}{\psi^{A}}\cr
{\delta}{\overline{\psi}}^{A}&=-i{\epsilon^{AB}}{\eta_{B}}
{\gamma_{\rho}}{\partial^{\rho}}{\phi}-{i\over 12}{\epsilon^{AB}}{\eta_{B}}
{\gamma_{\rho\kappa\lambda}}H^{\rho\kappa\lambda}\cr
{\delta}{\phi}&={\eta_{A}}{\overline{\psi}}^{A}+{\eta^{*}_{A}}{\psi^A}\cr}
\numbereq\name{\eqwdkf}
$$
In eq.(\eqwdkf) and elsewere in this section $\mu, \nu=1, \cdots ,6$.
The parameter $\eta_A, A=1, 2$ of the transformations
are spinors of $SO(6)$ and $\epsilon^{AB}=-\epsilon^{BA}, \epsilon^{12}
=1$. $\gamma_\mu$ are $\gamma$-matrices for six dimensions
and ${\gamma_{\mu\nu}}={\half}({\gamma_\mu}{\gamma_\nu}-
{\gamma_\nu}{\gamma_\mu}), {\gamma_{\rho\kappa\lambda}}={1\over
3!}({\gamma_{\rho}}{\gamma_{\kappa}}{\gamma_{\lambda}}-
{\gamma_{\kappa}}{\gamma_{\rho}}{\gamma_{\lambda}}+ cyclic)$.

As has been known for sometime, the closure of the supersymmetry algebra
requires the fields to obey the equations of motion
$$
H(x)-^* H(x)=0, \qquad {\gamma^{\alpha}}{\partial_{\alpha}}{\psi}(x)=0,
\qquad {\partial^2}{\phi}(x)=0.
\numbereq\name{\eqnakic}
$$
An action and canonical formulation of the first of these equations
is what we have obtained. The other two equations, viz., the Dirac
equation for $\psi$ and the d'Alembertian for $\phi$, can be derived
from the standard actions for these fields. The symplectic structures for
them are also standard.

\newsection Dimensional reduction.

The dimensional reduction of the self-dual field in six dimensions
where two of the dimensions form a torus is known to give a dual
symmetric version of electrodynamics [\verli]. The modular transformations
of the torus act as duality transformations on the Maxwell field. It is
interesting to see how this works out in terms of the action (\eqwaxops).

The five-dimensional spatial manifold is taken as $R^3 \times T^2$.
The torus $T^2$ is described as usual by the complex coordinate
$z={\sigma}_1 + {\tau}{\sigma}_2, 0 \leq {\sigma}_1, {\sigma}_2 \leq 1$,
$\tau$ being the modular parameter. The Coulomb Green's function
$-F(x, y)$ is given by
$$
-F(x, y)={\sum_n}{\int}{d^{3}k\over {(2{\pi})^3}}{ {
e^{2{\pi}i{\vec n}\cdot{({\vec\sigma}-{\vec\sigma}')}}}
\over {(Im{\tau})[{\vec k}^2+
4{\pi}^2g^{ij}{\vec n}_{i}{\vec n}_{j}]}}
\numbereq\name{\eqgalakte}
$$
where
$$
g_{ij}=\left (\matrix { 1&Re{\tau}\cr
                                     Re{\tau}&|{\tau}|^2\cr}\right).
\numbereq\name{\eqzevren}
$$
The ansatz for dimensional reduction is
$C={\beta}d{\sigma_2}-Bd{\sigma_1}$,
where $\beta, B$ are one-forms on $R^3$ and depend on the
coordinates $x_1, x_2, x_3$ of $R^3$. The action (\eqwaxops) can
now be worked out. We find
$$
S=-i{\int}d^{3}xd^{3}y~{(W_{i}(x){\dot{\overline W}}_{j}(y)
-{\overline W}_{i}(x)
{\dot W}_{j}(y))\over {Im{\tau}}}{\epsilon^{ijk}}{\partial_{\kappa}}g(x, y)
-{2\over Im{\tau}}{\int}d^{3}x~ W_{i}(x){\overline W}_{i}(x)
\numbereq\name{\eqgray}
$$
where
$$
W={\beta}+{\tau}B, \qquad g(x, y)={\int}{d^{3}k\over {(2{\pi})^3}}{
e^{i{\vec k} \cdot ({\vec x}-{\vec x}')}\over {{\vec k}^2}}.
\numbereq\name{\eqggud}
$$
( The first term in eq.(\eqwaxops) has an $\epsilon$-tensor. The integration
measure for this term is thus $d^{5}x~d^{5}y~{\sqrt g}$
where $g=(det g_{ij})$.
Apart from this observation , the integration over the coordinates
of $T^2$ is straightforward.)

It is easily checked that the modular transformation ${\tau} \to
-{1\over {\tau}}$ leaves $S$ unchanged apart from the replacement
${\beta} \to B, B \to {\beta}$. Integrating out $\beta$ in the functional
integral corresponding to $S$ of (\eqgray) will lead to the standard
Maxwell action, integrating out $B$ will give the dual action.
Consider the integration over $\beta$. Because of the condition
$\partial_{\mu}C_{\mu\nu}(x)=0$, we must have $\nabla \cdot{\beta}=0,
\nabla \cdot B=0$. The integration measure must take account of this.
Up to irrelevant constants, the measure for $\beta$-integration is thus
$$
d{\mu}(\beta)=\big [ d{\beta} \big ] {\delta} \big ( \nabla{\beta} \big )=
\big [ d{\beta}d{\varphi} \big ] e^{-i{\int}{\nabla_i}{\varphi}{\beta_i}}.
\numbereq\name{\eqanterson}
$$
We also introduce the potential $A_{i}(x)$ by
$$
A_{i}(x)=-{\int_{y}}{\epsilon_{ijk}}{\partial_{k}}g(x, y)~ B_{j}(y).
\numbereq\name{\eqpapadak}
$$
Integration of $e^{S}$ over ${\beta}, \varphi$ then leads to a reduced
action
$$
S=2{\int_{x}}Im{\tau} ~\big ( E^2- B^2)-4{\int_{x}}Re{\tau}~E \cdot B
\numbereq\name{\eqkambouris}
$$
where $E_{i}={\dot A}_i$. Writing $Im{\tau}={4{\pi}\over {g^2}}$ and
$Re{\tau}={{\theta}\over 2{\pi}}$ and rescaling the fields
$ ( E, B ) \to {1\over {\sqrt {16{\pi}}}}(E, B)$, the above equation becomes
$$
S={1\over 2g^2}{\int_{x}}\big ( E^2- B^2)-{{\theta}\over {8{\pi^2}}}
{\int_{x}}E\cdot B
\numbereq\name{\eqkmanris}
$$
which is the standard Maxwell action with a $\theta$-term.
Similarly integration over $B$ will give a dual description.

\newsection Acknowledgments.

VPN thanks R. Jackiw for discussions and N. Khuri for hospitality
at Rockefeller University.
This work was supported in part by the Department of Energy Contract
Number DE-FG02-91ER40651-TASKB and by the National Science Foundation Grant
number PHY-9322591.

\immediate\closeout1
\bigbreak\bigskip

\line{\twelvebf References. \hfil}
\nobreak\medskip\vskip\parskip

\input refs

\vfil\end

\bye